\RequirePackage{amsmath}
\documentclass[runningheads]{llncs}
\usepackage{graphicx}
\usepackage[loose]{subfigure} % subfigure
\usepackage{amsmath} % equation 
\usepackage{amssymb}
\usepackage[algo2e,lined,ruled]{algorithm2e} % algorithm layout
\usepackage{booktabs} % Pretty tables
\usepackage{helvet}
\usepackage{courier}
\usepackage{float}
\usepackage{url}
\usepackage{xspace}
\usepackage{comment}
\usepackage{alltt}
\usepackage{xcolor}
\usepackage{booktabs}
\usepackage{times}
\usepackage{wrapfig,fancyvrb}
\usepackage{color}
\usepackage{soul}

\SetAlFnt{\small}
\SetAlCapFnt{\small}
\SetAlCapNameFnt{\small}
\SetAlCapHSkip{0pt}
\IncMargin{-\parindent}

\begin{document}
\title{What if Process Predictions are not \\followed by Good Recommendations?\\(Technical Report)}
%
%\titlerunning{Predictions Fail Without Good Recommendations}
\author{Marcus Dees\inst{1,2}, Massimiliano de Leoni\inst{3},  \\Wil M.P. van der Aalst\inst{4,2} and Hajo A. Reijers\inst{5,2}}
\institute{Uitvoeringsinstituut Werknemersverzekeringen (UWV), The Netherlands
\and Eindhoven University of Technology, Eindhoven, The Netherlands 
\and University of Padova, Padova, Italy
\and RWTH Aachen University, Aachen, Germany
\and Utrecht University, Utrecht, The Netherlands \\
\email{marcus.dees@uwv.nl, deleoni@math.unipd.it, wvdaalst@pads.rwth-aachen.de, h.a.reijers@uu.nl}}

\maketitle              % typeset the header of the contribution
\begin{abstract}
Process-aware Recommender systems (PAR systems) are information systems that aim to monitor process executions, predict their outcome, and recommend effective interventions to reduce the risk of failure. This paper discusses monitoring, predicting, and recommending using a PAR system within a financial institute in the Netherlands to avoid faulty executions. Although predictions were based on the analysis of historical data, the most opportune intervention was selected on the basis of human judgment and subjective opinions. The results showed that, although the predictions of risky cases were relatively accurate, no reduction was observed in the number of faulty executions. We believe that this was caused by incorrect choices of interventions. Although a large body of research exists on monitoring and predicting based on facts recorded in historical data, research on fact-based interventions is relatively limited. This paper reports on lessons learned from the case study in finance and identifies the need to develop interventions based on insights from factual, historical data. 
\end{abstract}

\section{Introduction}
\label{sec:intro}
Process-aware Recommender systems (hereafter shortened as PAR systems) are a new breed of information systems. They aim to predict how the executions of process instances are going to evolve in the future, to determine those that have higher chances to not meet desired levels of performance (e.g., costs, deadlines, customer satisfaction). Consequently recommendations are provided on which effective contingency actions should be enacted to try to recover from risky executions. PAR systems are expert systems that run in the background and continuously monitor the execution of processes, predict their future, and, possibly, provide recommendations. Examples of PAR systems are discussed by Conforti et al. \cite{CONFORTI20151} and Schobel et al. \cite{SchobelR17}.
A substantial body of research exists on evaluating risks %\footnote{Note that, in fact, overspending is a special type of risk and, hence, cost evaluations are special types of risk evaluation}
, also known as process monitoring and prediction; see, e.g., the surveys by M\'{a}rquez-Chamorro et al. \cite{MarquezChamorro18} and by Teinemaa et al. \cite{TeinemaaDRM17}. %see, e.g., publications~\cite{CONFORTI20151,CONFORTI20132939,Fazzinga2018,Metzger17,PIKA201698,10.1007/978-3-319-59536-8_8,TeinemaaDMF16} and the survey~\cite{TeinemaaDRM17}.
Yet, as also indicated in~\cite{MarquezChamorro18}, \textit{``little attention has been given to providing recommendations''}.
In fact, it has often been overlooked how process participants would use these predictions to enact appropriate actions to recover from those executions that have a higher risk of causing problems. %One of the few research works that addresses this is~\cite{CONFORTI20151}.
It seems that process participants are tacitly assumed to take the ``right decision'' for the most appropriate corrective actions for each case. This also holds for approaches based on mitigation / flexibility ``by design''~\cite{6560815}. %or on operational research~\cite{Conforti12}.
Unfortunately, the assumption of selecting an effective corrective action 
is not always met in reality. When selecting an intervention, this is mainly done based on human judgment, which naturally relies on the subjective perception of the process instead of being based on objective facts. 

In particular, the PAR system should analyze the past process executions, and correlate alternative corrective actions with the likelihood of being effective; it should then recommend the actions that are most likely to decrease risks.
%We stress here the importance of building interventions on facts too: the system should reason on past process executions to correlate alternative corrective actions with the likelihood of being effective; it should then recommend the action that is most likely to decrease risks.
Otherwise, even if the monitor is able to drive the attention of process participants to those executions that actually require support, the recommender system is destined to ultimately fail. The positive occurrence of correctly monitoring a process and making an accurate prediction can be nullified by an improper recovery or intervention.
An organization will only profit from using a recommender system if the system is capable of making accurate decisions and the organization is capable of making effective decisions on the basis of this. Much attention is being paid to making accurate decisions, specifically to the proper use of data, measuring accuracy, etc.  \textit{In this work, we show that the analysis of making effective decisions is just as important. Both parts are essential ingredients of an overall solution.}

This paper reports on a field experiment that we conducted within UWV, a Dutch governmental agency. Among other things, UWV provides financial support to Dutch residents that lose their job and seek a new employment.
Several subjects (hereafter often referred to as customers) receive more unemployment benefits than the amount they are entitled to. Although this is eventually detected, it may take several months. Using the UWV's terminology, a \emph{reclamation} is created when this happens, i.e. a reclamation event is raised when a reclamation is detected.
To reclaim the amount of unlawfully provided support is very hard, time-consuming, and, often unsuccessful. In this context, an effective recommender system should be able to detect the customers who are more likely to get a reclamation and provide operational support to \textit{prevent} the provision of benefits without entitlement. %\footnote{Research at UWV has shown that the main causes for reclamations can be attributed to the customer making a mistake when informing UWV about income received next to their benefits.}

To follow up on this idea, we developed a predictor module that relies on machine-learning techniques to monitor and identify the subjects who are more likely to receive unlawful support. Next, various possible interventions to prevent reclamations were considered by UWV's stakeholders. The intervention that was selected to be tested in a field experiment consists of sending a specific email to the subjects who were suspected of being at higher risk.
\emph{The results show that risky customers were detected rather well, but no significant reduction of the number of reclamations was observed.}
This indicates that the intervention did not achieve the desired effect, which ultimately means that the action was not effective in preventing reclamations.
Our findings show the importance of conducting research not only on prediction but also on interventions. This is to ensure that the PAR system will indeed achieve the improvements that it aims at, hence creating process predictions that are followed by good recommendations.

The remainder of this paper is structured as follows. Section \ref{sec:UWV} introduces the situation faced at UWV and Section \ref{sec:research_method} shows which actions were taken, i.e., the building of a PAR and the execution of a field experiment. Section \ref{sec:results_achieved} discusses the results from the field experiment and Section \ref{sec:discussion} elaborates on the lessons learned from it. Section \ref{sec:conclusion} concludes the paper.

\section{Situation Faced -- The Unemployment-Benefits Process at UWV}
\label{sec:UWV}
UWV is the social security institute of the Netherlands and responsible for the implementation of a number of employee-related insurances. One of the processes that UWV executes is the unemployment benefits process. When residents in the Netherlands become unemployed, they need to file a request at UWV, which then decides if they are entitled to benefits. When requests are accepted, the customers receive monthly benefits until they find a new job or until the maximum period for their entitlement is reached.

\begin{figure}[t]
  \includegraphics[width=1.0\textwidth]{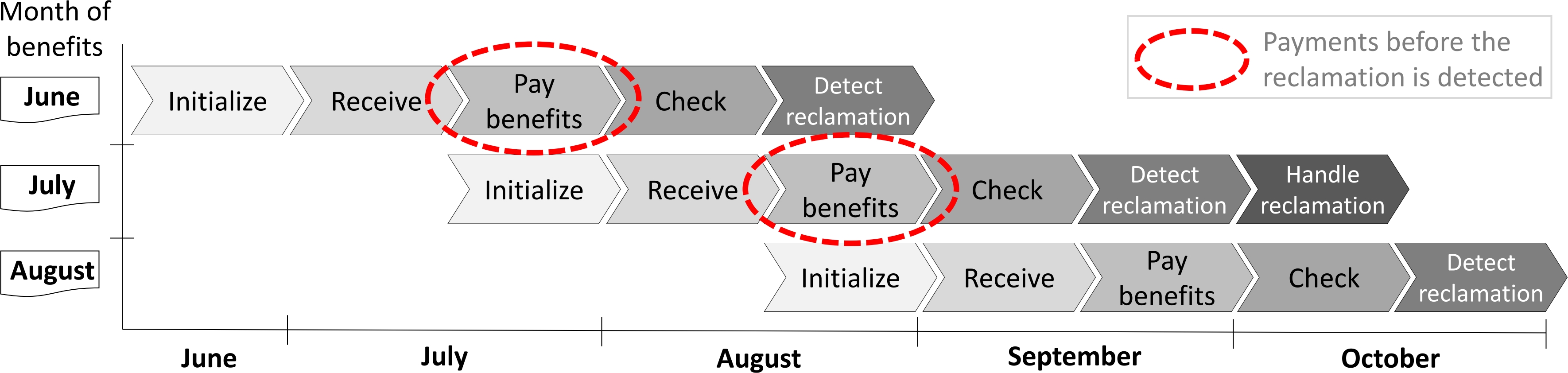}
  \caption{An example scenario of the potential activities that are related to the provision of the unemployment benefits for a customer for the months June, July and August (the year is irrelevant). Each row is related to the activities needed to handle an income form for the month of the benefits. Each benefits month takes several calendar months to be handled, e.g., the benefits for the month of June are handled from June until August.}
  \label{fig:unemployment_benefits_payment_process}
\end{figure}

The unemployment benefit payment process is bound by legal rules. Customers and employees of UWV are required to perform certain steps for each specific month (hereafter income month) in which customers have an entitlement.
Fig. \ref{fig:unemployment_benefits_payment_process} depicts a typical scenario of a customer who receives benefits, with the steps that are executed in each calendar month.
Before a customer receives a payment of benefits for an income month, an \textit{income form} has to be sent to UWV. Through this form customers specify whether or not they received any kind of income next to their benefits, and, if so, what amount. The benefits can be adjusted monthly as a function of any potential income, up to receiving no benefits if the income exceeds the amount of benefits to which the customer is entitled.
 
Fig.~\ref{fig:unemployment_benefits_payment_process} clearly shows that, in October, when the reclamation is handled, two months of unemployment benefits have already been paid, possibly erroneously.
Although  this seems a limited amount (usually a few hundred Euros) if one looks at a single customer, it should be realized that this needs to be multiplied by tens of thousands of customers in the same situation. 
UWV has on average 300,000 customers with unemployment benefits of whom each month on average 4\% get a reclamation.

The main cause for reclamations lie with customers not correctly filling in the amount of income earned next to their benefits on the income form.
%When a customer has income next to their benefits they have to fill in the earned amount. 
The correct amount can be obtained from the payslip. If the payslip is not yet received by the customer, they will have to fill in an estimate. However, even with a payslip it is not trivial to fill in the correct amount. The required amount is the \emph{social security wages}, which is not equal to the gross salary and also is not equal to the salary after taxes. %This is a reason for errors and cannot be changed because it is required by law.
An other reason for not correctly filling in the income form occurs when a customer is paid every 4-weeks, instead of every month. In this case there is one month each year with two 4-weekly payments. The second payment in the month is often forgotten. 
%A final example of a reason is related to the month in which holiday pay is received. This extra amount is also often not added to the regular amount. 
Apart from the reasons mentioned, there exist many more situations in which it can be hard to determine the correct amount.

Since the reclamations are caused by customers filling in income forms incorrectly, the only thing that UWV can do is to try to prevent customers from making mistakes filling in the income form. 
Unfortunately, targeting all customers with unemployment benefits every month to prevent reclamations can become very expensive.
Furthermore, UWV wants to limit communications to customers to only the necessary contact moments. Otherwise, communication fatigue can set in with the customers, causing important messages of UWV to have less impact with the customers.
Only targeting customers with a high chance of getting a reclamation reduces costs and should not influence the effectiveness of messages of UWV. Because of all these reasons, a recommender system that could effectively identify customers with a high risk of getting a reclamation would be really helpful for UWV. That recommender system needs to be able to target risky customers and propose opportune interventions.

\section{Action Taken -- Build PAR System and Execute Field Experiment}
\label{sec:research_method}

\begin{figure}[ht]
  \includegraphics[width=1\textwidth]{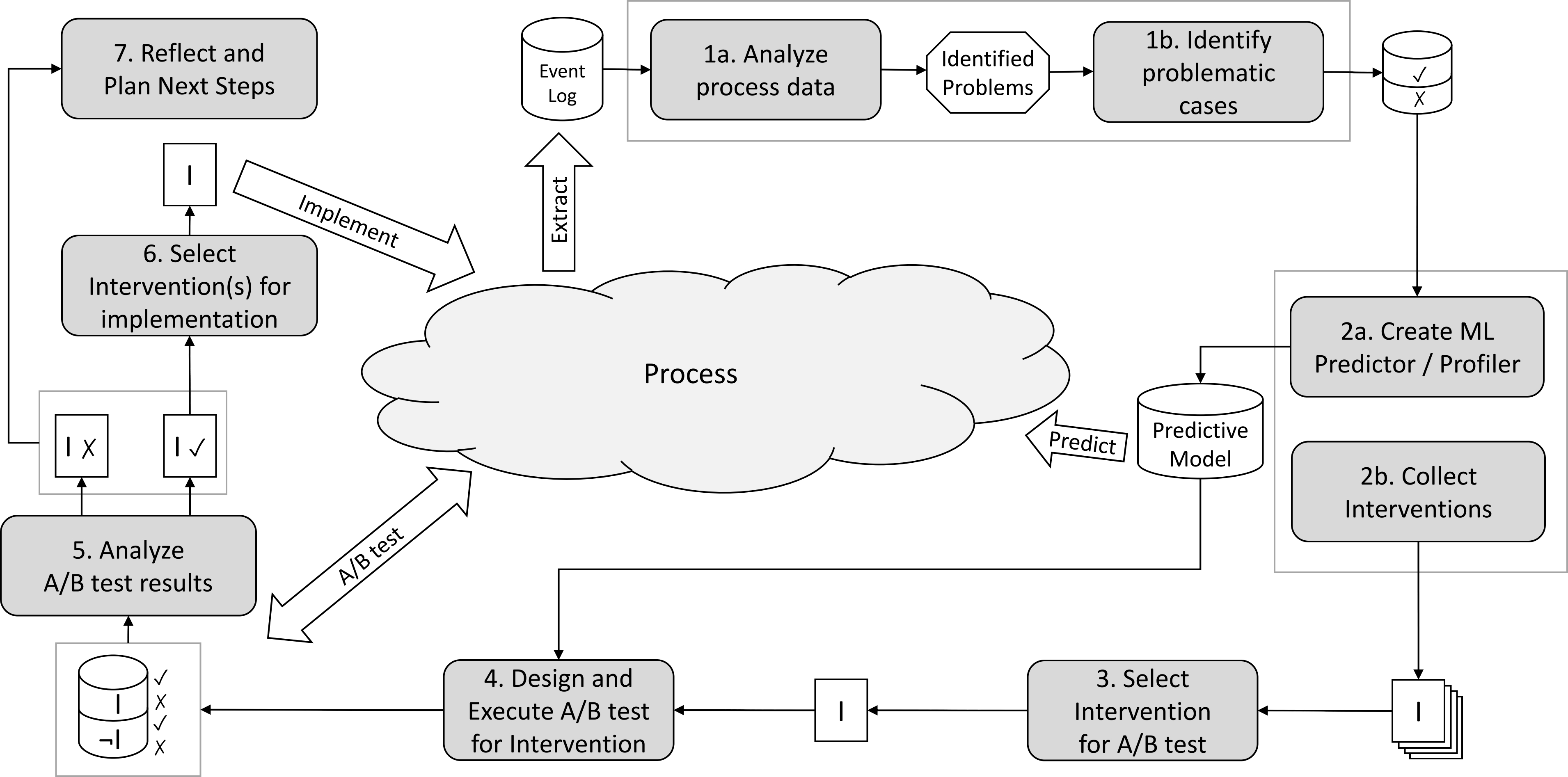}
  \caption{Overview of the steps that make up the research method. These steps correspond to one improvement cycle. The ``I'' is used as an abbreviation for ``Intervention''.}
  \label{fig:research_method}
\end{figure}

Our approach for the development and test of a PAR system for UWV is illustrated in Fig.~\ref{fig:research_method}. The first steps (Step 1a and 1b) of the approach are to analyze and identify the organizational issue. 
As described in Section \ref{sec:UWV} the organizational issue at UWV is related to reclamations.
%The first steps (Step 1a and 1b) of the approach are to analyze and identify the organizational issue. As described in Section \ref{sec:UWV} the organizational issue at UWV is related to reclamations.

The second step is to develop a recommender system, which consists of a predictor module (Step 2a) and a set of interventions (Step 2b). The predictor module is needed to identify the cases on which the interventions should be applied, namely the cases with the highest risk to have reclamations. Section \ref{subsec:prediction} describes the predictor module setup.

Together with the predictor module, an appropriate set of interventions needs to be selected. Interventions need to be determined in concert with stakeholders. Only by doing this together, interventions that have the support of the stakeholders can be identified. Support for the interventions is needed to also get support for the changes necessary to implement the interventions in the process. 

At UWV several possible interventions were put forward, from which one was chosen (Step 3). Only one intervention could be selected, due to the limited availability of resources at UWV to execute an experiment. Section \ref{subsec:mitigatingAction} elaborates on the collecting of interventions and selection of the intervention for the field experiment.

The next step is to design a field experiment (Step 4). The field experiment was set up as an \emph{A/B test} \cite{Kohavi2017}. 
In an A/B test, one or more interventions are tested under the same conditions, to find the alternative that best delivers the desired effect.
In our field experiment, risk type combined with the intervention can be tested in the natural setting of the process environment.
The objective of the field experiment is to determine the effect of applying an intervention for cases at a specific risk level, with respect to the specific process metrics of interest, i.e. whether or not a customer gets a reclamation. 
All other factors that can play a role in the field experiment are controlled, as far as this is possible in our business environment. 
Under these conditions, the field experiment will show if a causal relation exists between the intervention and the change in the values of the process metrics. 
Section \ref{subsec:setup_abtest} describes the setup for the UWV study.

The results of the field experiment are analyzed to determine if an effect can be detected from applying the intervention (Step 5). The desired effect is a reduction in the number of customers with a reclamation. Section \ref{subsec:intervention} and Section \ref{subsec:predmod} contain respectively the analysis of the intervention and the predictor module. If the intervention can be identified as having an effect, then both the direction of the effect, i.e. whether the intervention leads to better or worse performance, and the size of the effect need to be calculated from the data.
When an intervention has the desired effect, it can be selected to become a regular part of the process. The intervention then needs to be implemented in the process (Step 6). Interventions together with the predictor module from Step 2a, make up the PAR system. After the decision to implement an intervention it is necessary to update the predictor module of the PAR system. Changing the process also implies that the situation under which the predictions are made has changed. Some period of time after the change takes effect, needs to be reserved to gather a new set of historic process data on which the predictor module can be retrained.

The final step (Step 7) is the reflective phase in which the lessons learned from the execution of the approach are discussed. Within this research method, many choices need to be made. For example, which organizational issue will be tackled and which interventions will be tested. Prior to making a choice, the research participants should be aware of any assumptions or bias that could influence their choices. Section \ref{sec:discussion} contains the lessons learned for the UWV case.

\subsection{Building the Predictor Module}
\label{subsec:prediction}

The prediction is based on training a predictor component which uses historical data. This component was implemented as a stand-alone application in Python and leveraged the \emph{sci-kit learn} \cite{scikit-learn} library to access the data-mining functionality. 
For the UWV case, the historical data was extracted from the company's systems. It relates to the execution of every activity for more than 73,000 customers who concluded the reception of unemployment benefits in the period from July 2015 until July 2017. %Note that these customers, which were used to train the predictor, differ from the customers on whom the predictor was applied during the experiment. 

\begin{figure}[t]
\centering
  \includegraphics[width=1.0\textwidth]{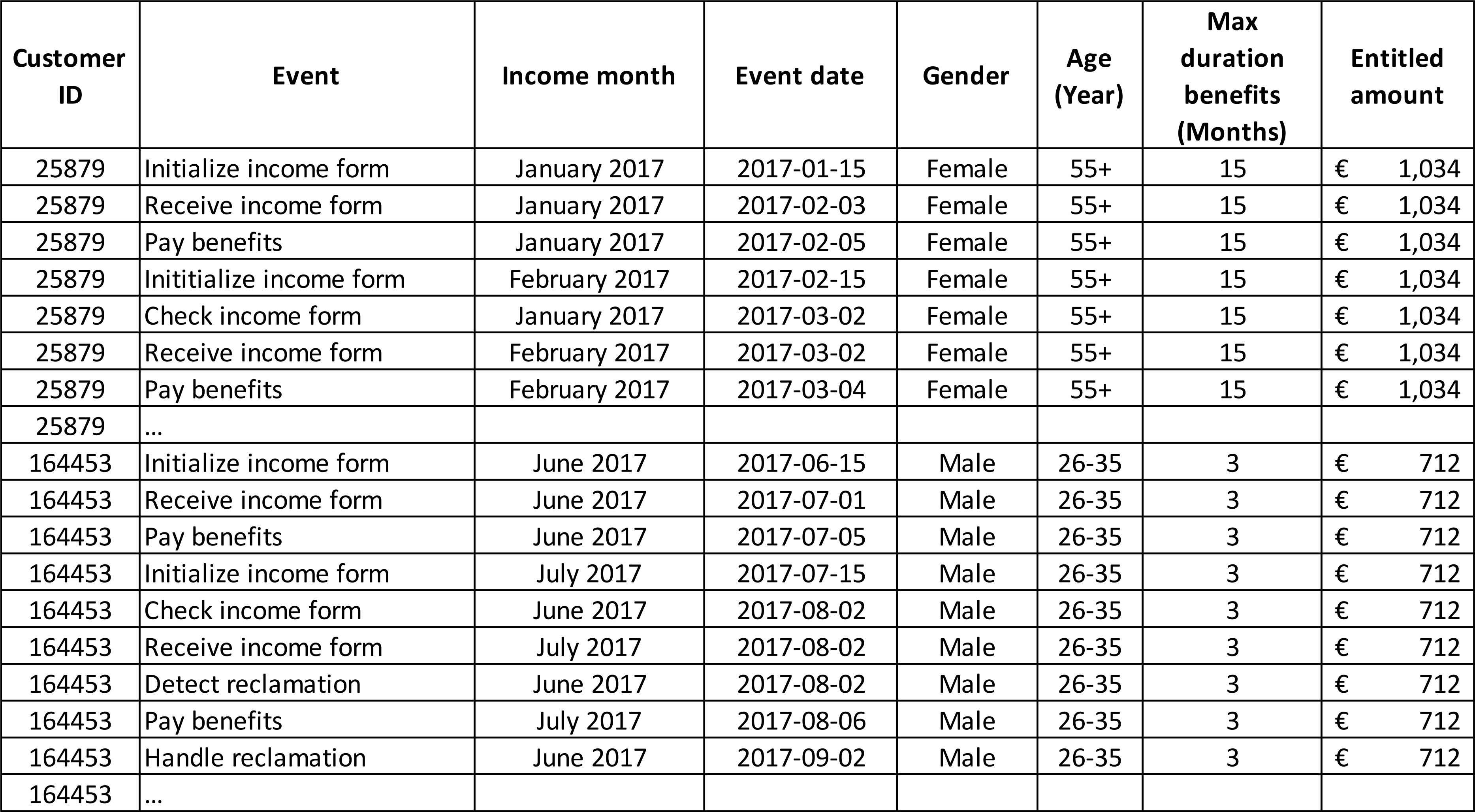}
  \setlength{\belowcaptionskip}{-12pt} 
  \caption{An example of an event-log fragment for two of UWV's customers. Each row refers to an event; events with the same \textit{Customer ID} are grouped into traces and ordered by \textit{Event date}.}
  \label{fig:eventLog}
\end{figure}

The collected information can be represented in a tabular form, of which an excerpt is presented in Fig.~\ref{fig:eventLog}. Each row in Fig.~\ref{fig:eventLog} corresponds to an event, namely the execution of an activity at a certain moment in time and refers to a customer with an identifier and other given characteristics.
The table forms a classical event log~\cite{procMiningBook}.
Events referring to the same customer can be grouped by customer id and ordered by timestamp, thus obtaining a trace.
For the UWV case study, the event log contained 5 million events (i.e. rows) for 73,153 customers, i.e.\ the event log contained 73,153 traces. Every trace refers to a complete execution of the process to provide benefits, which can end with finding a job or with reaching the end of the maximum benefit provision. A trace can contain zero, one, or more reclamation events.

The classifier of the prediction module is trained using the traces of the UWV's event log as input. Similarly to what is proposed in~\cite{TeinemaaDMF16,TeinemaaDRM17}, each trace $\sigma=\langle e_1, e_2, ..., e_m \rangle$ is encoded into a vector of variables that contains:
\begin{enumerate}
  \item The number of executions of each process' activity in $\sigma$ (one numeric variable per activity);
  \item the number of months for which the unemployment benefit can be maximally given (one numeric variable);
  \item the duration of the process execution in terms of number of months, i.e.\ the number of months existing between $e_1$ and $e_m$ (one numeric variable);
  \item customer characteristics, such as age, gender, and marital status;
  \item properties of the employment that triggered the unemployment benefits like the sector, the type of contract, working pattern and the reason for the dismissal;
  \item the presence / absence of a reclamation (one Boolean variable) at the end of $\sigma$.
\end{enumerate}

\begin{figure}[t]
  \includegraphics[width=1\textwidth]{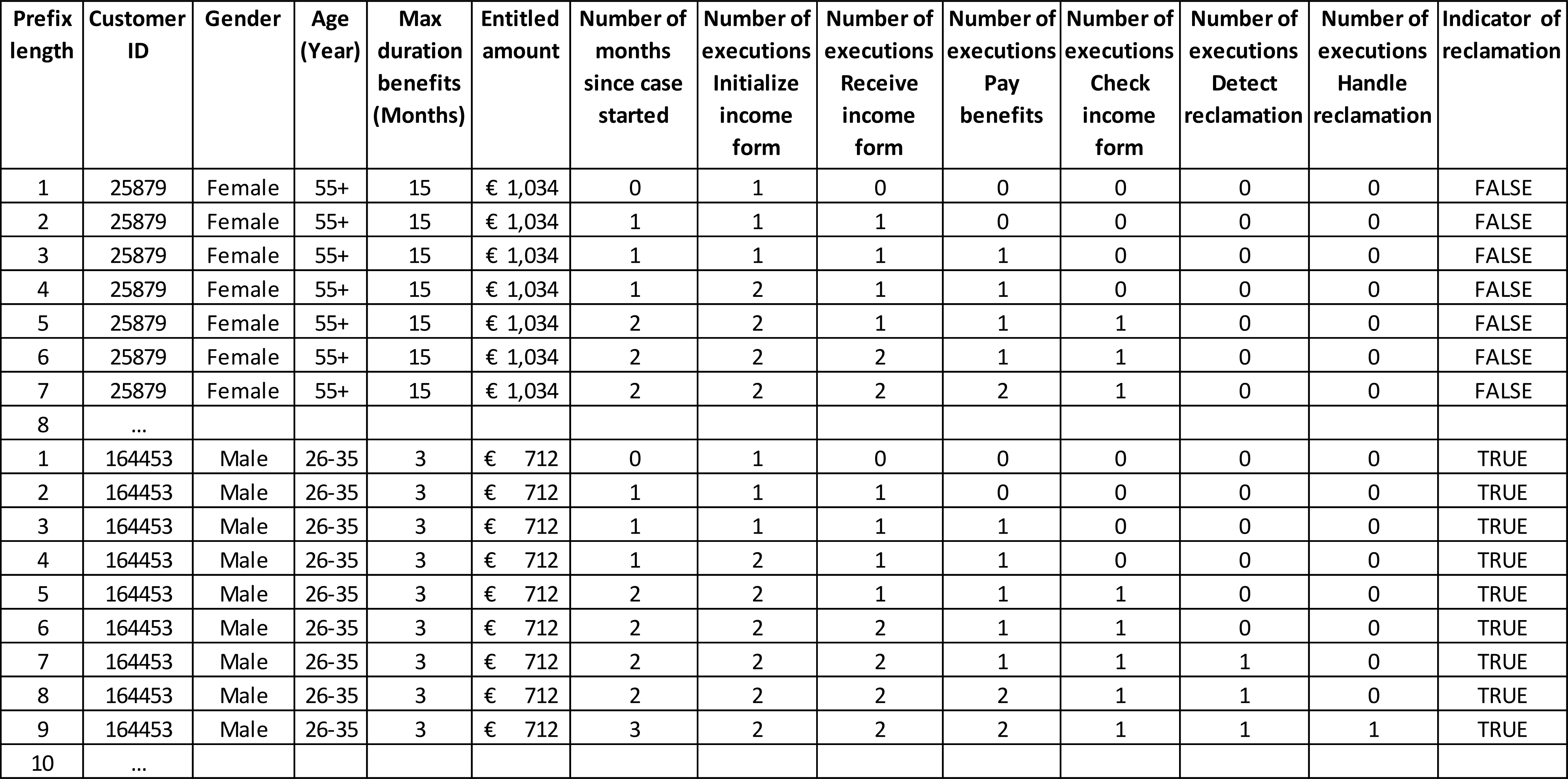}
  \setlength{\belowcaptionskip}{-12pt} 
  \caption{Example of vectors that are used as instances to train the predictor. These vectors correspond to the excerpt of the event log in Fig.~\ref{fig:eventLog}}
  \label{fig:preprocessed_dataset}
\end{figure}

\noindent Since we want to predict running cases and the event log records completed cases, we need to consider prefixes of running process instances. Namely, if a trace $\sigma=\langle e_1, e_2, ..., e_m \rangle$ is composed by $m$ events, we build $m$ prefixes: $\langle e_1\rangle,\langle e_1, e_2, \rangle,\ldots$, $\langle e_1, e_2, ..., e_m \rangle$. These prefixes are treated as running cases, with the notable difference that the eventual, actual outcome is known. In this way, the prefixes are a suitable input for training the predictor. Fig. \ref{fig:preprocessed_dataset} shows the set of training vectors that are generated for the two traces depicted in Fig.~\ref{fig:eventLog}. 
Each prefix is encoded as mentioned above, which includes the Boolean variable about the presence / absence of a reclamation at the end of the whole $\sigma$, called \textit{Indication of reclamation}. This variable is used as the dependent variable, where the others are used as independent variables to correlate with the dependent.
As an example, the first instance refers to the execution for customer $25879$ when only the first activity \textit{Initialize the Income Form} was considered; the second is about the same customers when the first and second activity were accounted. 

Teinemaa et al.\ illustrate that several data-mining predictors can be used to predict the dependent variable that encodes the KPI outcome~\cite{TeinemaaDRM17}, ranging, e.g., from Decision Tree, Random Forest, and Support Vector Machine till Generalized Boosted Regression Models, Logistic Regression, and ADA Boost.
For our experiments, we decided to opt for Logistic Regression and ADA Boost because they provide a predicting model that allows one to analyze which vector's components are most heavily affecting the prediction (e.g., the beta value of Logistic Regression). 
As discussed in Section~\ref{sec:UWV}, the frequency with which activities are executed for each customer is of the order of once a month. Therefore, it is not worthwhile predicting and recommending more than once a month. So only prefixes referring to entire months are retained; in other words, we train the predictor using the prefix $\langle e_1, e_2, ..., e_i \rangle$ of trace $\sigma=\langle e_1, e_2, ..., e_m \rangle$ if $e_{i+1}$ belongs to the month that follows that of $e_i$. E.g., looking at Fig.~\ref{fig:preprocessed_dataset} for customer $25879$, we train on the prefixes of length 1, 4, and 7, because these represent the last prefixes before the next month starts. 

The techniques based on Logistic Regression and ADA Boost were tuned through hyper-parameter optimization \cite{ClaesenM15}. To this end, the UWV's event log was split into a training set with 80\% of the traces and a test set with 20\% of the traces. The models were learned through a 5-fold cross validation on the training set, using different configurations of the algorithm's parameters. The models trained with different parameter configurations were tested on the second set with 20\% of the traces and ranked using 
the area under the ROC curve (shortened as AUC)~\cite{ROC}. AUC was chosen because it is the most suitable criterion in case of unbalanced classes: the number of customers with reclamation is just around 4\% of the total number.
When performing hyper-parameter optimization, we also tested two alternatives, as advised by Teinemaa et al.~\cite{TeinemaaDMF16,TeinemaaDRM17}. The first alternative is to train a single predictor with the vectors referring to all prefixes referring to whole months (see discussion above about the prefixes retained). The second alternative was to cluster the vectors according to the length of the corresponding prefixes in months, and to assign each vector cluster to a different predictor. Therefore, one predictor was trained of the vector of prefixes spanning over one month, one predictor with those spanning over two months, etc. The outcome of the hyper-parameter optimization was that the second alternative generally led to higher AUC values in combination with the ADA Boost technique.

\subsection{Collecting and Selecting the Interventions}
\label{subsec:mitigatingAction}

After three brainstorm sessions, with 15 employees and 2 team managers of UWV, the choice of the intervention was made by the stakeholders.
As mentioned earlier, the choice of intervention was based on the experience and expectations of the stakeholders. The aim of the intervention is to prevent customers from incorrectly filling the income form. More specifically, to prevent the customer from filling in the wrong amount.
The sessions initially put forward three potential types of interventions. The types are defined based on the actors that are involved in the intervention (the customer, the UWV employee, or the last employer):

\begin{enumerate}
  \item the customer is supported in advance on how to fill the income form;
  \item the UWV employee verifies the information provided by the customer in the income form, and, if necessary, corrects it after contacting the customer;
  \item the last employer of the UWV customer is asked to supply relevant information more quickly, so as to be able to promptly verify the truthfulness of the information provided by the customer in the income form; 
\end{enumerate}
An intervention can only be executed once a month, namely between two income forms for two consecutive months.
In the final brainstorming session, out of the three intervention types, the stakeholders finally opted for option 1 in the list above, i.e. supporting the customer to correctly fill the income form.
Stakeholders stated that, according to their experience, their support with filling the form helps customers reduce the chance of incurring in reclamations. As mentioned earlier, only one specific intervention was selected for the experiment, due to the limited availability of resources at UWV. 

The selected intervention entails pro-actively informing the customer about specific topics regarding the income form, which frequently lead to an incorrect amount.
These topics relate to the definition of social security wages, financial unemployment and receiving 4-weekly payments instead of monthly payments. The UWV employees indicated that they found that most mistakes were made regarding these topics. 

Next to deciding the action, the medium through which the customer would be informed, had to be determined. The options were: a physical letter, an email, or a phone call by the UWV employee. In the spirit of keeping costs low, it was decided to send the support information by email. An editorial employee of UWV designed the exact phrasing. The email contained hyperlinks to web pages of the UWV website to allow customers to obtain more insights into the support information provided in the email itself. The customers to whom the email was sent were not informed about the fact that they were targeted because they were expected to have a higher risk of getting a reclamation.
A tool used by UWV to send emails to large numbers of customers at the same time provided functionality to check whether the email was received by the recipient, namely without a bounce, as well as whether the email is opened by the customer's email client application. 
Since the timing of sending the message can influence the success of the action, it was decided to send it on the day preceding the last working day of the calendar month in which the predictor module marked the customer as risky. This ensured that the message could potentially be read by the customer before filling in the income form for the subsequent month. 

\subsection{Design and Execution of the Field Experiment}
\label{subsec:setup_abtest}
The experiment aims to determine whether or not the use of the PAR system would reduce the number of reclamations in the way it had been designed in terms of prediction and intervention. Specifically, we first determined the number and the nature of the customers who were monitored. Then, the involved customers were split into two groups: on one group the PAR system was applied, i.e. the experimental group, the second group was handled without the PAR system, i.e. the control group.

We conducted the experiment with 86,850 cases, who were handled by the Amsterdam branch of UWV. These were customers currently receiving benefits, and they are different from the 73,153 cases who were used to train the predictor module. Out of the 86,850 cases, 35,812 were part of the experimental group.
The experiment ran from August 2017 until October 2017. On 30 August 2017, 28 September 2017 and 30 October 2017 the intervention of sending an email was executed. 
The predictor was used to compute the probability of having a reclamation for the 35,812 cases of the experimental group. The probability was higher than 0.8 for 6,747 cases, and the intervention was executed for those cases.

\section{Results Achieved}
\label{sec:results_achieved}

The intervention did not have a preventive effect even though the risk was predicted reasonably accurate. Sections \ref{subsec:intervention} and \ref{subsec:predmod} describe the details of the results achieved.

\subsection{The Intervention Did Not Have a Preventive Effect}
\label{subsec:intervention}

\begin{figure}[t]
\begin{center}
  \includegraphics[width=0.95\textwidth]{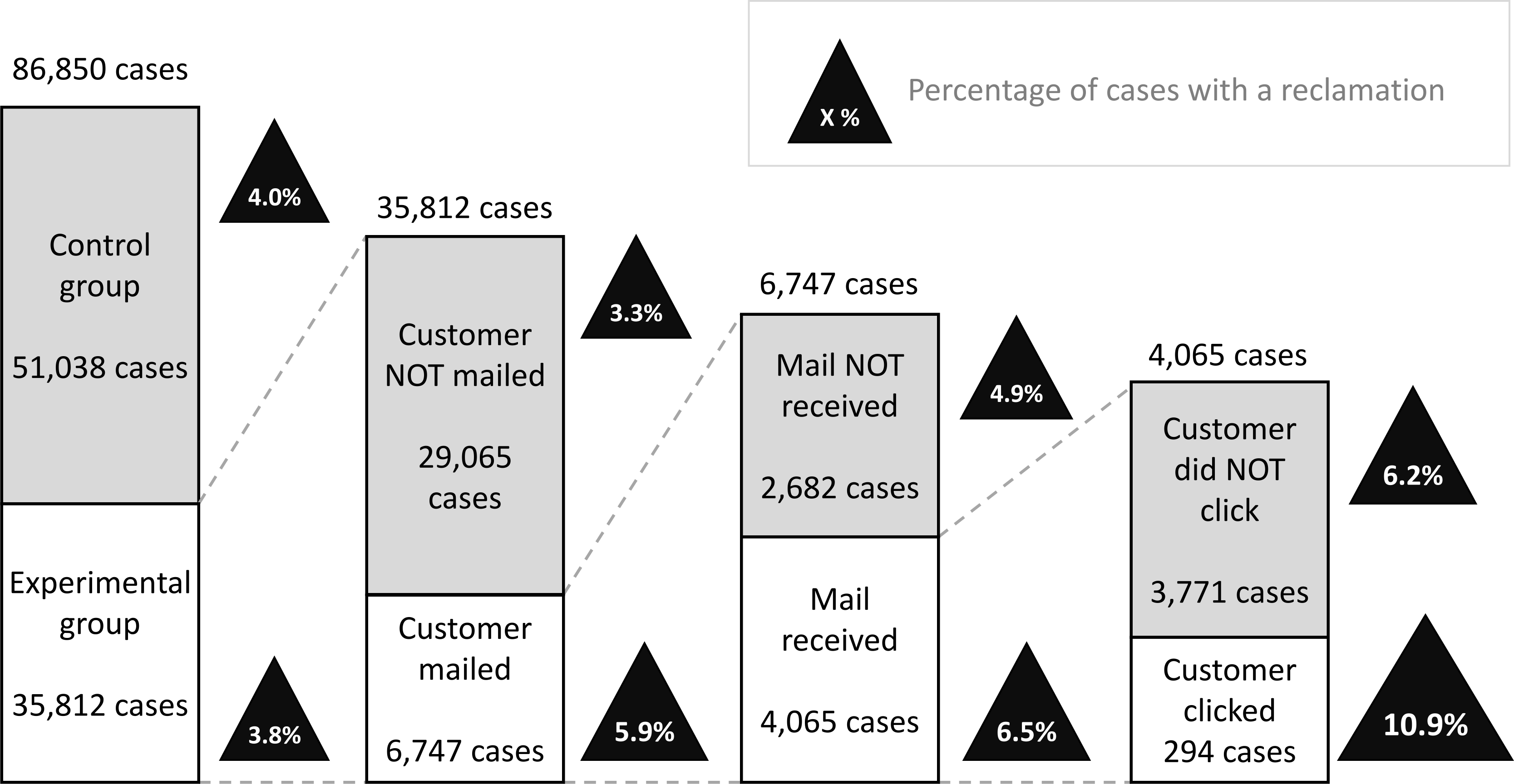}
  \caption{The number of cases and percentage of cases having a reclamation for all groups. The results show that risky customers are identified, but the intervention does not really help.}
  \label{fig:results_group}
\end{center}
\end{figure}
% \begin{figure}[t!]
%   \includegraphics[width=0.92\textwidth]{PredMon_Breakdown_of_results_C05_2.jpg}
%   \caption{The number of cases and percentage of cases having a reclamation for all groups. The results show that risky customers are identified, but the intervention does not really help.}
%   \label{fig:results_group}
% \end{figure}

Fig.~\ref{fig:results_group} shows the results of the field experiment, where the black triangles illustrate the percentage of reclamations observed in each group. The triangles at the left-most stacked bar show that the number of reclamations did not significantly decrease when the system was used, i.e. from 4.0\% without using the system to 3.8\% while using the system. The effectiveness of the system as a whole is therefore 0.2\%.

The second bar from the left shows how the PAR system was used for the customers: 6,747 cases were deemed risky and were e-mailed. Out of these 6,747 cases, 4,065 received the emails with the links to access further information. The other 2,682 cases did not receive the email. As mentioned in Section~\ref{subsec:mitigatingAction} the tool that UWV uses for sending bulk email can detect whether an email is received and is opened, i.e. there was no bounce. Since there were almost no bounces, the cases that did not receive the email, actually did not open the message in their email client.
From the customers who have received the email, only 294 actually clicked on the links and accessed the UWV's web site. Remarkably, among the customers who clicked the link, 10.9\% of those had a reclamation in the subsequent month: this percentage is more than 2.5 times the average. Also, it is around 1.7 times of the frequency among the customers who received the email but did not click the links.

\begin{figure}[ht]
\begin{center}
  \includegraphics[width=0.92\textwidth]{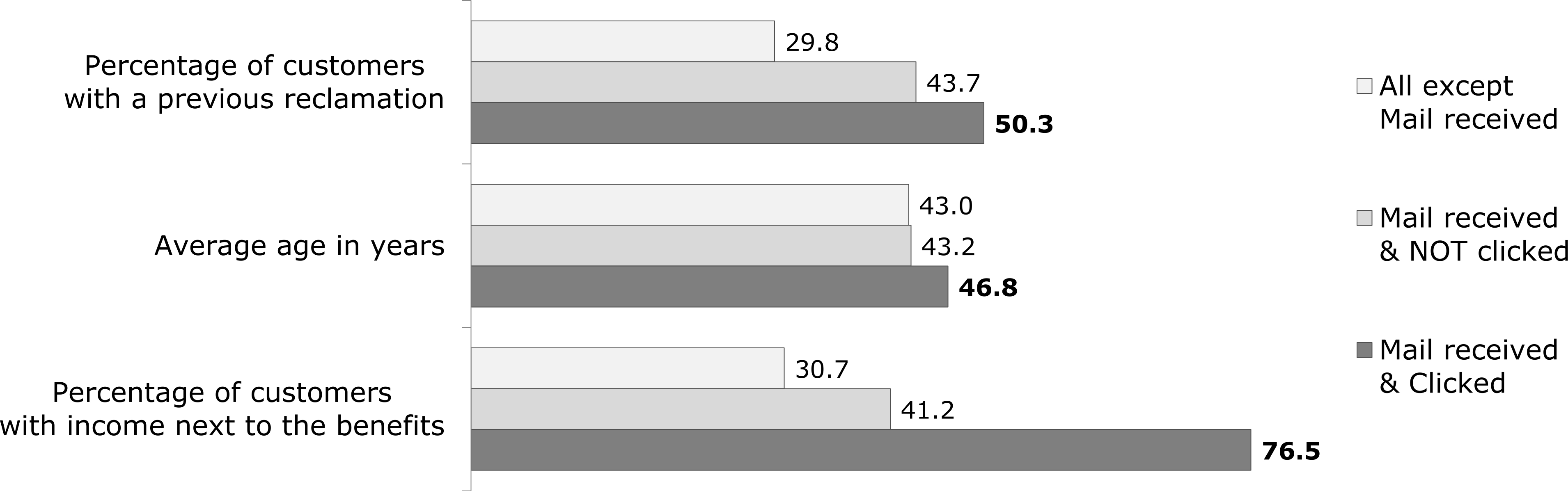}
  \setlength{\belowcaptionskip}{-12pt}
  \caption{Comparison of characteristics of customers who did not receive the email, those who received it but did not click the link and who accessed UWV's web site through the email's link. }
  \label{fig:dif}
\end{center}
\end{figure}

We conducted a comparative analysis among the customers who did not receive the email, those who received it but did not click the links and, finally, those who reached the web site.
The results of the comparative analysis are shown in Fig.~\ref{fig:dif}. The results indicate that 76.5\% of the customers who clicked the email's links had an income next to the benefits. 
Recall that it is possible to receive benefits even when one is employed: this is the situation when the income is reduced and the customer receives benefits for the difference.
It is a reasonable result:  mistakes are more frequent when filling the income form is more complex (e.g., when there is some income, indeed).
Additional distinguishing features of the customers who clicked on the email's link are that 50.3\% of these customers have had a previous reclamation, as well as that these customers are on average 3.5 years older, which is a statistically significant difference.

% \begin{figure}[b!]
% \begin{center}
%   \includegraphics[width=0.92\textwidth]{PredMon_Difference_between_experimental_groups_C05.jpg}
%   \caption{Comparison of the characteristics of the customers who did not receive the email, those who received it but did not click the link and who accessed the UWV's web site through the email's link. }
%   \label{fig:dif}
% \end{center}
% \end{figure}
%
The results even seem to suggest that emailing appears counterproductive or, at least, that there was a positive correlation between exploring the additional information provided and being involved in a reclamation in the subsequent month. To a smaller extent, if compared with the average, a higher frequency of reclamations is observed among the customers who received the email but did not click the links: 6.2\% of reclamations versus a mean of 3.8-4\%.
A discussion on the possible reasons for these results can be found in Section~\ref{sec:discussion}. However, it is clear that the intervention did not achieve the intended goal.

\subsection{The Risk Was Predicted Reasonably Accurate}
\label{subsec:predmod}

As already mentioned in Section \ref{sec:intro} and Section \ref{subsec:intervention}, the analysis shows the experiment did not lead to an improvement. To understand the cause, we analyzed whether this was caused by inaccurate predictions or an ineffective intervention or both. 
In this section, we analyze the actual quality of the predictor module. We use the so-called \textit{cumulative lift curve} \cite{DBLP:conf/kdd/LingL98} to assess the prediction model. This measure is chosen because of the imbalance in the data as advised in \cite{DBLP:conf/kdd/LingL98}.
% NOTE: the difference between ROC and Cum Lift Curve is that in the ROC on the x-axis the cumulative False Positive Rate is plotted (FP / TN + FP), while the Cum Lift Curve shows (TP + FP / all), e.g. the cumulative positive rate.
As mentioned before in Section \ref{sec:UWV}, only 4\% of the customers are eventually involved in reclamations.
In cases of unbalanced data sets (such as between customers with and those without reclamations), precision and recall are less suitable to assess the quality of predictors. Furthermore, because of the low cost of the intervention of sending an email, the presence of \textit{false negatives}, here meaning those customers with undetected reclamations \textit{during the subsequent month}, is much more severe than \textit{false positives}, i.e. customers who are wrongly detected as going to have reclamations \textit{during the subsequent month}. 

\begin{figure}[ht]
\begin{center}
  \includegraphics[width=0.70\textwidth]{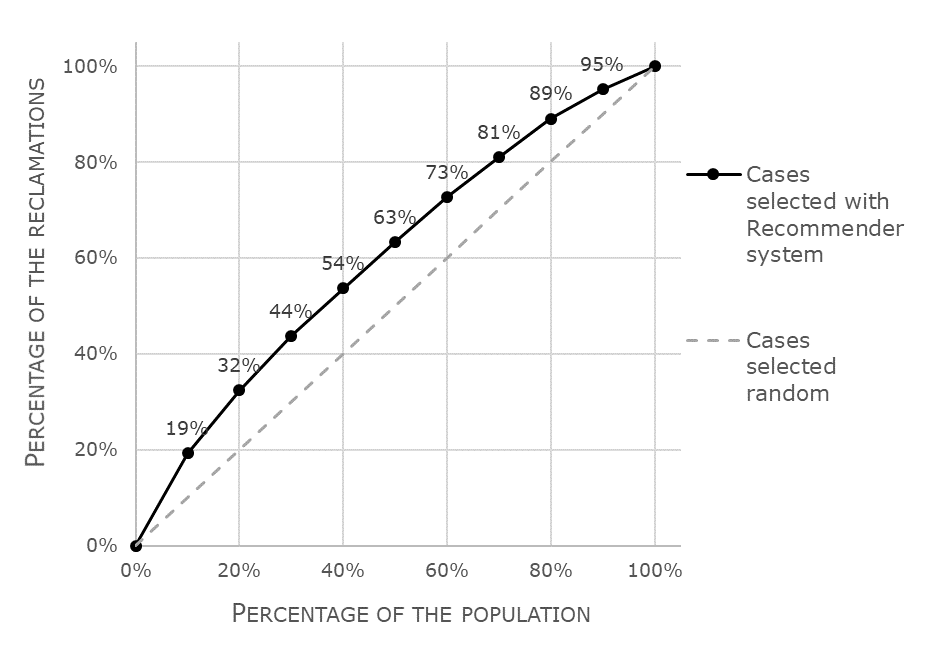}
  \setlength{\belowcaptionskip}{-12pt}
  \caption{The cumulative lift curve shows that using the recommender system leads to a better selection of cases than using a random selection of cases.}
  \label{fig:cum_lift_curve}
\end{center}
\end{figure}

% \begin{figure}[b!]
% \begin{center}
%   \includegraphics[width=0.75\textwidth]{PredMon_CumLiftCurve_C02.png}
%   \caption{The cumulative lift curve shows that using the recommender system leads to a better selection of cases than using a random selection of cases.}
%   \label{fig:cum_lift_curve}
% \end{center}
% \end{figure}

Fig.~\ref{fig:cum_lift_curve} shows the curve for the case study at UWV.  
The rationale is that, within a set of $x\%$ of randomly selected customers, one expects to observe $x\%$ of the total number of reclamations. This trend is shown as a dotted line in Fig.~\ref{fig:cum_lift_curve}. In our case, the predictions are better than random. For example, the 10\% of customers with the highest risk of having a reclamation accounted for 19\% of all reclamations, which is roughly twice as what can be expected in a random sample. 

%A second way to show the prediction accuracy can be derived from Fig.~\ref{fig:results_group}. The percentage of reclamations in the group of customers not mailed (as part of the experimental group) is 3.3\%, which is lower than the expected historical average of 4.0\%. This means that customers who had a higher probability of reclamation than 4.0\% were indeed included in the group of customers that were mailed (i.e. as a result of the prediction). This is another indication of the accuracy of the prediction.

In summary, although the prediction technique can certainly be improved, a considerable prediction effectiveness can be observed (cf.\ Section~\ref{subsec:prediction}). However, as mentioned in Section~\ref{subsec:intervention}, the system as a whole did not bring a significant improvement. This leads us to conclude that the lack of a significant effect should be mostly caused by the ineffectiveness of the intervention. In Section~\ref{sec:discussion}, we discuss this in more detail.

\section{Lessons Learned}
\label{sec:discussion}

The experiment proved to be unsuccessful. On the positive side, the predictions were reasonably accurate. However, the intervention to send an email to high risk customers did not lead to a reduction in the number of reclamations.
There even was a group of customers who had twice as many reclamations as the average population. Section~\ref{subsec:why_not_work} elaborates on the reasons why the intervention did not work. Section~\ref{subsec:what_done_different} focuses on the lesson learned, delineating how the research methodology needs to be updated. 

\subsection{Why Did the Intervention Not Work?}
\label{subsec:why_not_work}
One of the reasons why the intervention was not successful might be related to the wrong timing of sending the email. A different moment within the month could have been more appropriate. However, this does not explain why of the 6,747 cases selected only 294 acted on the email by clicking the links.
Other reasons may be that the customers might have found the email message unclear or that the links in the email body pointed to confusing information on the UWV website. In the group of 294 cases who clicked the links and who took notice of this information a reclamation actually occurred 2.5 times as much. 

Also, the communication channel could be part of the cause. Sending the message by letter, or by actively calling the customer might have worked better. In fact, when discussing reasons of the failure of the experiment, we heard several comments from different stakeholders that they did not expect the failure because \textit{``after speaking to a customer about how to fill in the income form, almost no mistakes are made by that customer''} (quoted from a stakeholder). This illustrates how the subjective feelings can be far from objective facts.

\subsection{What Should be Done Differently Next Time?}
\label{subsec:what_done_different}
We certainly learned that the A/B testing is really beneficial to assess the effectiveness of interventions. The involvement of stakeholders and other process participants, including, e.g., the UWV's customers, is beneficial towards achieving the goal.
However, the results did not achieve the expected results. We learned a number of lessons to adjust our approach that we will put in place for the next round of the experiments:
\begin{enumerate}
  \item Creating a predictor module requires the selection of independent features as inputs to build the predictive model. 
  From the reflection and the analysis of the reasons that caused the failure of an intervention, one can derive interesting insights into new features that should be incorporated when training the predictor. For instance, the features presented in Fig.~\ref{fig:dif} can be used to train a better predictor for the UWV case. 
  These features could be, e.g., a boolean feature whether a customer has income next to the benefits.
  \item The insights discussed in the previous point, which can be derived from the analysis, can also be useful to putting forward potential interventions. 
  For instance, an intervention could be to perform a manual check of the income form when a customer has had a reclamation in the previous month. This intervention example is derived from the feature representing the number of executions of \emph{Detect Reclamation} as discussed in Section \ref{subsec:intervention}.
  \item Before the selection of the interventions for the A/B test (Step 3 in Fig.~\ref{fig:research_method}), they need to be pre-assessed. The intervention used in our experiment is about providing information to the customers concerning specific topics related to filling the income form. In fact, before running the experiments, we could have already checked on the historical event data whether the reclamations were on average fewer when providing information and support to fill the income form. If this would had been observed, we could prevent ourselves from running experiments destined to fail.
  \item Since a control group was compared with another group on which the system was employed and the comparison is measured end-to-end, it is impossible to state the reason of the failure of the intervention, beyond just observing it. For instance, we
  should have used questionnaires to assess the reasons of the failure: the customers that received the email should have been asked why they did not click on the links or, even if clicked, still were mistaken. Clearly, questionnaires are not applicable for all kinds of interventions. Different methods also have to be envisaged to acquire the information needed to analyze the ineffectiveness of an intervention.
  \item It is unlikely that the methodology in Section~\ref{sec:research_method} already provided satisfactory results because of the methodology needs to be iterated in multiple cycles. In fact, this finding is compliant with the principle of \emph{Action Research}, which is based on idea of continuous improvement cycles ~\cite{CG@ECRM03,Rowell2017}.
  \item The point above highlights the importance of having interaction cycles. However, one cycle took a few months to be carried out. This is certainly inefficient: the whole cycle needs to be repeated at high speed and multiple interventions need to be tested at each cycle. Furthermore, if an intervention is clearly ineffective, the corresponding testing needs to be stopped without waiting for the cycle to end.
\end{enumerate}
All the lessons learned share one leitmotif: \textit{accurate predictions 
are crucial, but their effect is nullified if it is not matched by effective recommendations, and effective recommendations must be based on evidence from historical and/or experimental data}. 

\begin{figure}[t]
\begin{center}
  \includegraphics[width=1\textwidth]{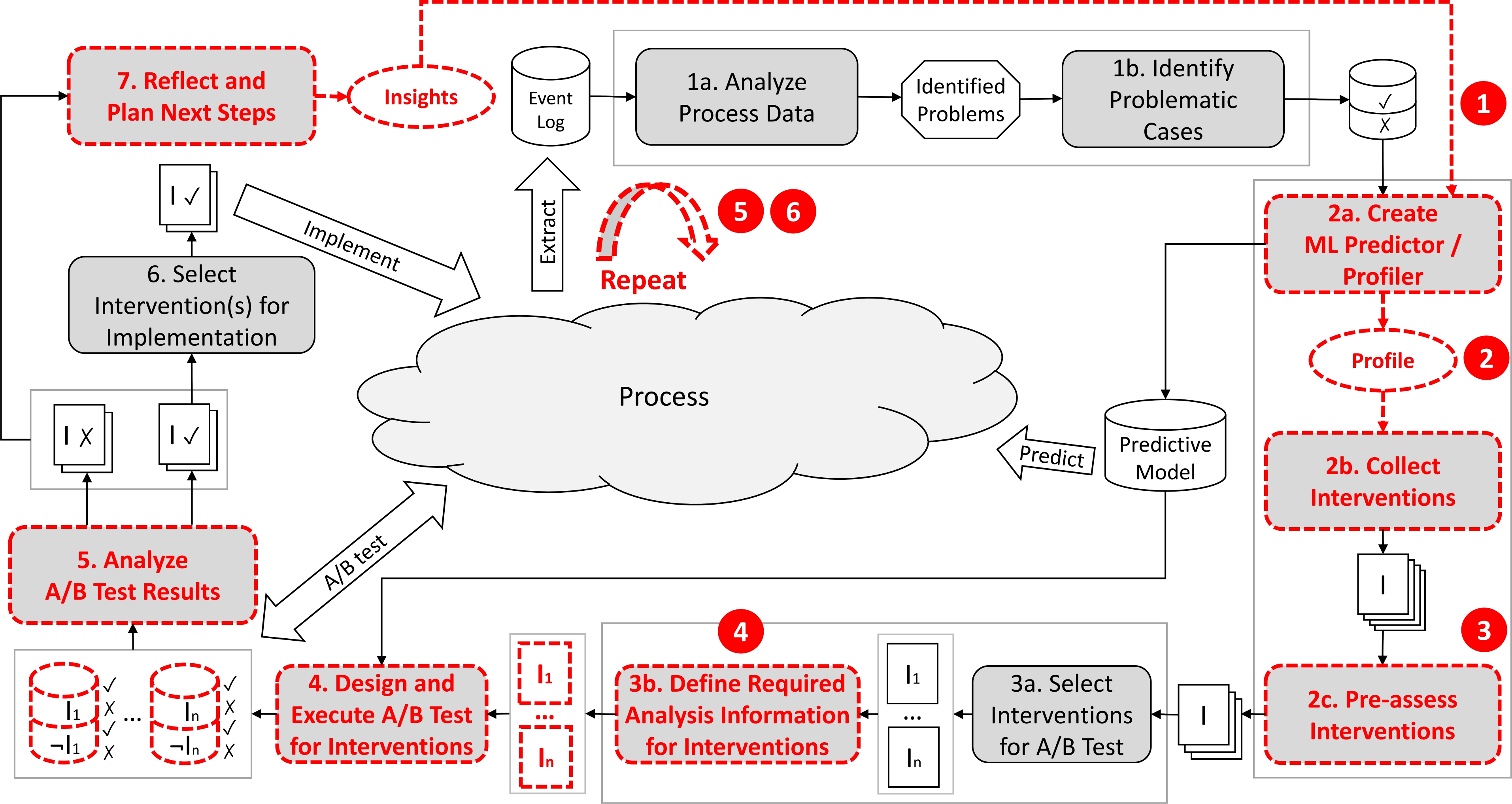}
  \setlength{\belowcaptionskip}{-14pt} 
  \caption{Overview of the steps that make up the \emph{updated research method}. These steps correspond to one improvement cycle and are repeated in every cycle. The ``I'' is used as abbreviation for ``Intervention''. The components that are changed relative to Fig.~\ref{fig:research_method} have red dashed lines.}
  \label{fig:par_system}
\end{center}
\end{figure}

\noindent In light of this, the methodology introduced in Section~\ref{sec:research_method} needs to be adjusted; the resulting new methodology is shown in Fig.~\ref{fig:par_system}.  The changes relative to Fig.~\ref{fig:research_method} are shown in red dashed lines.
To show how the lessons learned have impacted the original methodology, the items of the previous list are mapped on Fig.~\ref{fig:par_system} as numbers within a red circle.

The impact of adapting the research method according to the lessons learned is not limited to the identified components. For example, the second lesson has impact on the collection of interventions. Generating interventions in a data-driven manner is added to the stakeholder-driven approach. The third lesson adds the new pre-assessment step to the approach (Step 2c). The result of this step is the deselection of interventions collected in Step 2b. The fourth lesson introduces Step 3b, in which the information needed to understand the (in)effectiveness of an intervention is defined. Defining this information has an impact on the design of the A/B Test and the analysis of the results in Step 5. For example when questionnaires need to be deployed.

Lesson 5 and 6, i.e. repeat the cycle, speed it up and use multiple interventions, are not linked to one specific step. These lessons have impact on the whole approach. 
Since the updated approach is more elaborate than the original approach it will require more effort to execute one cycle of this method, let alone multiple cycles with multiple interventions at a high speed. Systematic support needs to be developed for all of the steps of the updated research methodology to allow for a smooth execution.

\section{Conclusion}
\label{sec:conclusion}
When building a Process-aware Recommender system, both the predictor module and the recommender parts of the system must be effective in order for the whole system to be effective. In our case, the predictor module was accurate enough. However, the intervention did not have the desired effect. The lessons learned from the field experiment are translated into an updated research method. The updated approach asks for high speed iterations with multiple interventions. Systematic support will be needed for each step of the approach to meet these requirements.

As future work, we plan to improve the predictor module to achieve better predictions by using different techniques and leveraging on contextual information about the customer and its history, e.g., the presence of some monetary income next to the benefits is strongly causally related to reclamations.
As described, we want to use evidence from the process executions, and insights from building the predictor module, to select interventions to be tested in a new experiment.

Orthogonally to a new field experiment, we aim to devise a new technique that adaptively finds the best intervention based on the specific case. Different cases might require different interventions, and the choice of the best intervention should be automatically derived from the historical facts recorded in the system's event logs. In other words, the system will rely on machine-learning techniques that (1) reason on past executions to find the interventions that have generally been more effective in the specific cases, and (2) recommend accordingly.  

\bibliographystyle{splncs}
\bibliography{paper}

\end{document}